# Testing match-3 video games with Deep Reinforcement Learning

Nicholas Napolitano*

**Abstract** Testing a video game is a critical step for the production process and requires a great effort in terms of time and resources spent. Some software houses are trying to use the artificial intelligence to reduce the need of human resources using systems able to replace a human agent. We study the possibility to use the Deep Reinforcement Learning to automate the testing process in match-3 video games and suggest approaching the problem in the framework of a Dueling Deep Q-Network paradigm. We test this kind of network on the Jelly Juice game, a match-3 video game developed by the redBit Games. The network extracts the essential information from the game environment and infers the next move. We compare the results with the random player performance, finding that the network shows a highest success rate. The results are in most cases similar with those obtained by real users, and the network also succeeds in learning over time the different features that distinguish the game levels and adapts its strategy to the increasing difficulties.

**Keywords** Machine Learning · Video Games · Deep Reinforcement Learning · Match-3

## 1 Introduction & Related Work

The methods of machine learning can be used to improve the computer software, for example automating the process of code and output testing [2, 11]. The evaluation of the quality of video games can take a long time because it requires a large amount of tests [5] and it would be therefore desirable to have a synthetic system to simulate the behavior of the human players [6]. The inputs of the video games are mainly visual and can be easily decoded by means of convolutional neural networks [10]. The same networks can be also a candidate tool to infer the rules of the game by observing the behavior of the real players [7].

The problem is significant for its applications and has been studied for the match-3 game *Candy Crush*, a game which falls into the NP-Hard and NP-Complete categories for the case of unbounded boards [15]. The game used to work in this paper has the same structure and complexity of the one above.

The possibility of predicting the difficulty of a match-3 games in terms of average human success rate (AHSR) was explored by means of *Monte Carlo Tree Search* (MCTS). However this approach is a long-lasting move predictor for a video game [12]. In order to improve the move selection algorithm for each iteration, in our method we recognize the depth level reached in the tree of the possible moves.

The MCTS strategy has been also coupled with a Deep Learning system to predict the difficulty of a game to make a MCTS agent faster than, for example, a random player [13]. Moreover, other Reinforcement Learning approaches have already been used to train agents to play match-3 games using a Model-Free approach on a fixed board with definite specifications [9]. Instead, in our work we create a model which can be used in many forms of board and objectives.

In this paper we analyze the performance of a neural network which learns how to play a match-3 video game, whose goal is to align on a grid three objects of the same type swapping two adjacent elements. We set up a network able to learn how to play and progressively improve its performance testing the model on the

*Corresponding Author: Nicholas Napolitano
Dipartimento di Ingegneria, Università Roma Tre, Rome, Italy, 00146
Tel.: +39-3933260379
E-mail: nic.napolitano1@stud.uniroma3.it

Jelly Juice video game[1] developed by the redBit Games software house[2].

We find that the Deep Reinforcement Learning is an effective strategy for a synthetic system to play match-3 games improving skill and performance. Comparing the Dueling Deep Q-Network and the random strategy with the human players behavior, we also find that the former is more suitable to simulate the real players behavior. Although a random strategy can appear to be more efficient because of its simplicity and speed, a Dueling Deep Q-Network allows to compare the behavior of a synthetic player with that of a human player. The remainder of the paper is organized as follows. We discuss the Match-3 games in Sect. 2 and the model in Sect. 3. Then, we present main results in Sect. 4 and finally our conclusions are given in Sect. 5.

## 2 The Match-3 games

The board of a match-3 game is a square lattice $\Lambda \subset \mathbb{R}^2$ on which are arranged the variables $a_k \in A, k \in \Lambda$ modeling the tiles shown in Fig.1, the *blockers* which hinder the actions of the players, and the *jokers*. A configuration at time $t$ is a set of values $a(t) = \{a_k(t) \mid k \in \Lambda\}$ and a move is a swap of the values of two neighboring variables. If the move does not lead to any match-3 (horizontal or vertical alignment of three or more variables with the same value) the previous configuration is restored. When a match-3 is achieved the new configuration replaces the previous one. In this case the move is considered a valid move because it decreases the value of the game counter. The simplest goal of this kind of game is the collection of a given number of match-3 within the available number of moves.

We consider lattices with $m$ rows, and $n$ columns, where $3 \leq m, n \leq 9$. The variables $a_k$ are uniquely *One Hot Encoded* as vectors of $0, 1$ of length $|A|$, so that $a_k \in \{0, 1\}^{|A|}$ with the constraint $||a_k|| = 1$. The space of the configurations of the game is $C = A^{|\Lambda|}$ and the number of the possible moves can be easily established by the following statement.

**Theorem 1** *For the $m \times n$ lattice $\Lambda$, the number $N$ of the possible moves is*

$$N = 2(2mn - m - n) \quad (1)$$

*Proof* The $N_a$ variables on the corners have only 2 possible swaps, the horizontal and the vertical one. The $N_b$ variables on the edges can do 3 swaps: vertical, horizontal and one more on one of these directions. The $N_i$

[1] https://www.redbitgames.com/portfolio/jelly-juice/
[2] https://www.redbitgames.com/

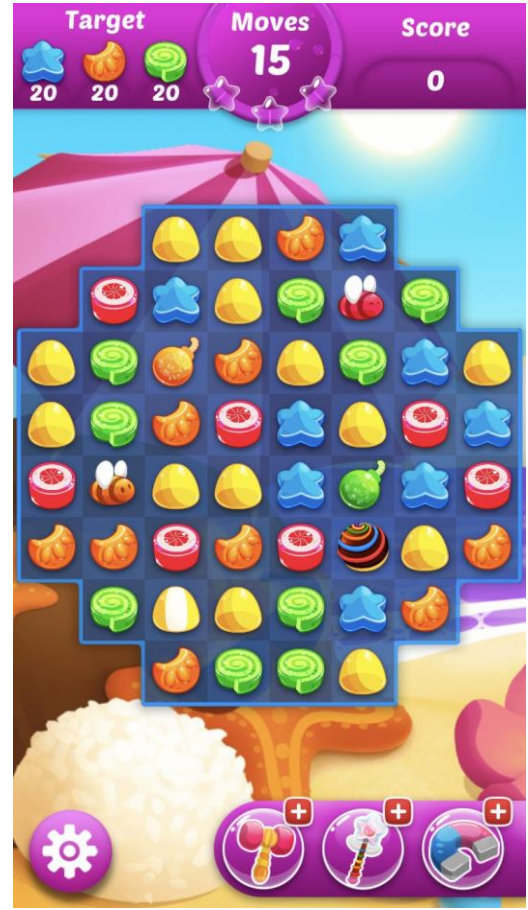

Fig. 1: Game Map of Jelly Juice

non-edge tiles have 4 possible swaps, 2 horizontal and 2 vertical.

The total number of moves is therefore

$$N = 4N_i + 3N_b + 2N_a \quad (2)$$

Since $N_a = 4$, $N_b$ is the lattice perimeter without the vertices, given by $2(n-2) + 2(m-2)$, and $N_i$ is the internal area given by $(n-2)(m-2)$, replacing these values in Eq.(2) we have $N = 2(2mn - m - n)$

We notice that $N$ is always even and for the $n \times n$ square lattice the moves are $N = 4n(n-1) < 4|\Lambda|$. A move $m$ is the swap of the values of two neighboring $a_i$ and is represented by a couple of tiles.

$$m = (a_i, a_j) \in A \times A \equiv M \quad (3)$$

For the small number of moves the exhaustive search could be plausible, but our goal is to find a system that simulates the human behavior. The comprehensive search is useful for testing the game performance, but not the behavior of the players. The game strategy is given in terms of the probability $P\left(m \mid a(t)\right)$ to swap the values of the neighboring variables ($a_i, a_j$) conditioned by the game configuration $a(t)$. The simplest possible player is the *random player* where the probabilities do not change with time, while the *perfect player* is the algorithm choosing the best move for each iteration $t$.

A Human-like AI can be trained from the random player to converge towards the behavior of a perfect player (also called *super-human AI*). A machine of this type falls into the Human-Like AI category [16]. At the beginning all the tiles have the same probability to be swapped. To modify these probabilities, we need to compute the utility score of a Reinforcement Learning model.

## 3 The Artificial Player

To model the artificial player, we take a $9 \times 9$ lattice $\Lambda$, and $a_k \in A = \{1, 2, \ldots, 6\}$. The set $C$ of the possible configurations has cardinality $|C| = 6^{9 \times 9}$. With these assumptions we set up a neural network with four layers, the categorical-cross entropy for the loss function and the standard Adam optimizer [1]. The first three layers interact by the *ReLU* activation function, but to compute the probabilities we use for the last layer the *softmax* function. The training of the model on a set of effective moves for very few epochs showed that the average accuracy reaches a plateau at 80%, with an average loss of 0.54 and test accuracy of 10%.

We also tested variants of this basic associative network: VGG16, BrainNet, ResNet, DenseNet[3] and MobileNetV3 [8], obtaining similar results. The increasing of the number of training data from the initial 10,000 to 60,000 did not improve the performance, but increasing progressively the number of epochs from 30 to 10,000, a new local minimum was found, characterized by the 91% of training accuracy and 0.26 of loss. During the first training the neural network was stuck in a local minimum about the 90% of the computation time (80 hours). To estimate the training time, we computed the Mean Time Between Failures [3], the average time to wait for the jump from one metastable state to another. The training was run on 60,000 inputs, 10,000 epochs and the clock of the processor was 280,800 Hz. With a network of 705 neurons, 81 for the input layer, 100 for the second layer, 200 for the third and 324 for the output, we had metastability time of 152.8 hours, about six days and a half.

The Jelly Juice game gives configurations with at least one possible match-3, and we modeled the moves by means of the two *tap* and *swap* operators to compute the new configuration $a^j \in C$, with $a^j \neq a$. From 10,000 different configurations we selected 319,648 valid moves, and trained the neural network for 100 epochs observing that the loss decreased regularly by 0.0001 every five epochs and gave a final values lower than 0.2 with accuracy about 13%. A further 1000 epochs training led to an improvement of only 2% for the accuracy.

[3] https://keras.io/applications/

The trained network was the starting point of a Reinforcement Learning model with the reward given by the Hamming distance between $a(t)$ and $a(t+1)$

$$d(a(t), a(t+1)) = \sum_{k \in \Lambda} (1 - \delta_{a_k(t), a_k(t+1)}) \qquad (4)$$

measuring the number of replacements necessary to transform $a(t)$ into $a(t+1)$, where $\delta_{ij}$ is 1 when $i = j$ and 0 otherwise. In fact, we consider the move that gen- erates a configuration very different from the previous one to be more effective because was produced by many cascading match-3.

Exploiting the Q-Learning algorithms, the optimal move prediction improved, but not the accuracy. For the states, $a \in C$ and the actions $m \in M$, the set of swap actions is given by the table of correspondences

$$Q : C \times M \to [0, 1] \qquad (5)$$

which gives the probabilities associated to the swaps and updated by the Bellman Equation [14]. We also introduced the *Experience Replay Memory* (ERM), a data structure whose elements are the tuples $(a, m, r, a^j)$, where $a$ is the starting state, $m$ the action, $r$ the reward attributed to the action and $a^j$ the target state.

For each iteration a fixed length set of samples of the Replay Memory was chosen, the prediction of the next state was performed and the corresponding Q-Value updated depending on the reward, on the depth of descent into the tree of the moves and on the discount factor $\gamma \in [0, 1)$. We notice that the non-valid moves influence the game dynamics, the non-valid moves with non-zero probabilities slow down the syn- thetic player. Therefore, we give non zero probability only to the moves that can lead to a change of config- uration to achieve of the greatest possible success rate and obtaining the highest possible score.

## 4 Results

After some attempts to reduce the number of layers, the optimal network turned out to be formed by two Dense Neural Networks run by a Deep Reinforcement Learning algorithm. The network architecture, that we call JellyGym, is shown in Fig.2 and is given by one input layer with 82 neurons, 81 for the board and one for the current objective, a hidden layer fully-connected with 100 neurons and one more hidden layer fully-connected with 200 neurons. The output layer has 324 neurons, one for each swap (also including the forbidden swaps).

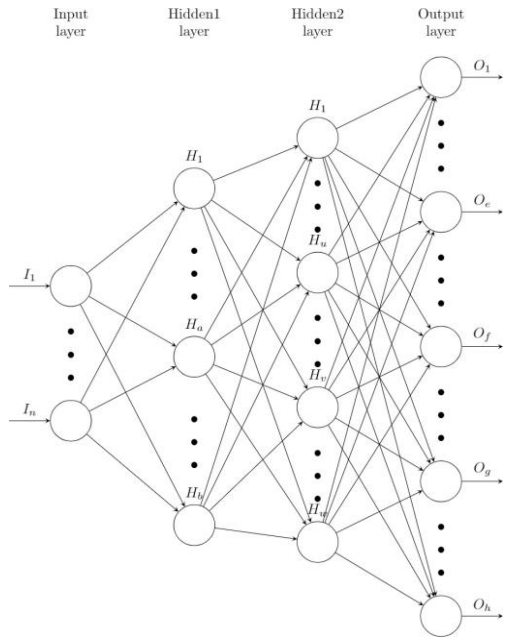

Fig. 2: JellyGym's Neural Network Architecture

The *ReLU* activation function connects the input layer and the first hidden layer and between the first and the second hidden layer and the *Softmax* activation function transforms the second hidden layer output into the set of probabilities. The learning rate was $\alpha = 0.001$, the loss function the *Mean Squared Error* and the optimizer given by the *Adam* function. A *batch normalization* was applied to each layer, improving the learning rate, simplifying the initialization of the weights and making faster the training [4]. The size of the network was taylored on the maximum possible size of the board. For the smaller boards we used a *zero padding* to standardize all the boards to a unique format filling it with zeros (empty tiles) to achieve the standard size. The output was the set of the probabilities of the moves, including also the non-valid ones.

The behavior of the **Dueling Deep Q-Network** (DDQN) was simulated by means of two copies of the network, one for the prediction of the next move starting from the current state and trained at each iteration (single move made during the game) and the second one to predict the best move starting from the next state (*Dueling Network* or *Oracle*) with the weights of the first one at each episode (single game played). We also used an arbitrary length success case history (*Experience Replay memory*) for each iteration whose elements are given by the tuple (*State, action, reward, NextState*). The utility value is a positive reward if the current move allowed to get closer to the current goal and a negative penalty otherwise. We used an increasing discount factor depending on the moves left in the counter.

In summary the training was given by the following steps.

The first one involved the use of Supervised Learning, providing immediate feedback with respect to the performed action. If the utility value was positive, a gradient descent of 1000 epochs started with the pair formed by the current configuration together with the index of the move, where $-1$ replaced the zeros and the utility value replaced the 1 to make it more effective.

The second step was based on the DDQN algorithm, with a set of sub-cases of the Replay Memory for each tuple:

– The prediction of the next state using the Main Network and the Oracle.
– The update of the Q-Value relating to the current state and the move predicted by the Main Network using the Bellman Equation. The utility value was discounted depending on the number of the past moves:

$$Q^{new}(s_t, a_t) \leftarrow \underbrace{(1-\alpha) \cdot Q(s_t, a_t)}_{\text{old value}} + \underbrace{\alpha}_{\text{learning rate}} \cdot \bigg( \underbrace{r_t}_{\text{reward}} + \underbrace{\gamma}_{\text{discount}} \cdot \underbrace{\overbrace{\max_a Q(s_{t+1}, a)}^{\text{learned value}}}_{\text{optimal value}} \bigg)$$

(6)

– A gradient descent of 100 epochs was run on the couple formed by the current state and the updated Q-values.

The third step involved a further Supervised Learning block. From the current configuration we got the feasible moves cycling on all the moves to penalize the non-valid ones before predicting the next one. We computed the probabilities of the possible moves, storing into a list the valid ones and normalized the probabilities to exclude the non-valid moves. Finally we predicted the move.

The time of a single move prediction was approximately 60 seconds on a *NC6_Promo* Standard Virtual Machine with 6 vCPUs and 56 GB of Memory and 1 NVIDIA TESLA v100 GPU. We compared the performances of four possible players: the *Random Player* choosing the move with fixed probability, the *JellyGym* which chooses the best move, the *Smart Player* which chooses with the same probability one of the moves getting the best type of Jolly and finally the *Real Users*. The game levels are characterized by different tasks. The performances for each type of player are summarized in the table Tab.1. For all the levels JellyGym achieved a success rate higher than that obtained by the random player (in the advanced levels a random strat- egy has never allowed to overcome them). In addition, JellyGym achieved in most levels results comparable to that of real users. The evolution of the success rate as functions of the played matches is shown in Fig.3. We observe that JellyGym requires a given number of training games to be able to adapt to the characteristics of the current environment. Then, the value grows gradually while the system has "understood" how to solve the level and in the rest of the matches it undertakes to progressively increase its performance. We found that 150 matches were enough to show this phenomenon for all the levels except for the level 505, whose complexity was too high.

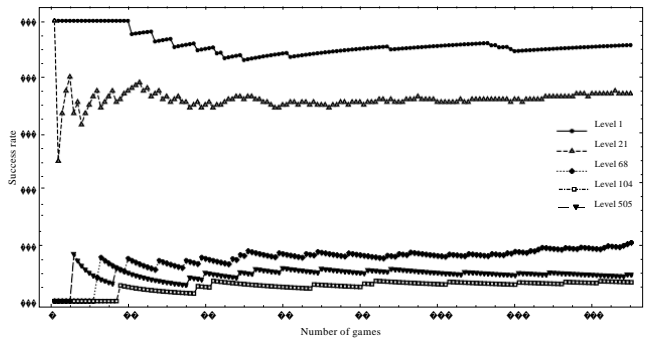

Fig. 3: Success rate of the Jelly Gym system for different levels

| Level | Matches Played | Random Player | JellyGym | Smart Player | Users |
|---|---|---|---|---|---|
| 1 | 150 | 65% | 91.3% | 94% | 96.58% |
| 21 | 150 | 54% | 74% | 95.9% | 94.7% |
| 68 | 150 | 0 % | 20.67 % | 32.3 % | 24.24% |
| 104 | 150 | 0 % | 6.67 % | 13.5% | 8.2 % |
| 505 | 150 | 0 % | 9.33 % | 10.52 % | 13.05 % |

Table 1: Success rate Jelly Juice levels

## 5 Conclusions

The test of a video game is a central operation for the development process, but is expensive in terms of time and work. The artificial intelligence methods can help to automate and optimize the process. We studied a possible optimal strategy to set up a synthetic player. The results show that the Deep Reinforcement Learning is a promising approach because the neural network improves its performance with experience. Different approaches experimented, such as Supervised Learning through a dataset of configurations and moves, as well as different Reinforcement Learning architectures did not give equally appreciable results.

Our results are consistent with those of other researches aimed to examine artificial intelligence systems to solve match-3 problems [12] [13] [9]. The system based on the Dueling Deep Q-Network shows a better performance than a random player in every level and exhibits a behavior like those of human users in almost every match level. In the more advanced levels a greater number of games is necessary to make the model training effective, but the data show that even for more complex levels the automaton responds satisfactorily.

Some interesting aspects have been identified that can be explored in future. It would be interesting to test the same network architecture on different types of games to verify the generality of the algorithm, checking whether the learning and the training on Jelly Juice gives an efficient system to play also Candy Crush or more general match-3 games.

**Acknowledgements** We acknowledge Gabriele Achler, Francesco Comi and Massimo Guareschi of the RedBit Games company for making available the Jelly Juice codes, as well as granting work space, funds and computational resources to work on the project. We also thank Giuseppe Sansonetti and Alessandro Micarelli of the Dipartimento di Ingegneria Università Roma Tre for their support.

## Conflict of interest

The authors declare that they have no conflict of interest.

## References

1. Abadi, M., Barham, P., Chen, J., Chen, Z., Davis, A., Dean, J., Devin, M., Ghemawat, S., Irving, G., Isard, M., Kudlur, M., Levenberg, J., Monga, R., Moore, S., Murray, D.G., Steiner, B., Tucker, P., Vasudevan, V., Warden, P., Wicke, M., Yu, Y., Zheng, X.: Tensorflow: A system for large-scale machine learning. In: 12th USENIX Symposium on Operating Systems Design and Implementation (OSDI 16), pp. 265 – 283 (2016). URL https://www.usenix.org/system/files/conference/osdi16/osdi16-abadi.pdf
2. Ammar, H., Abdelmoez, W., Hamdi, M.: Software engineering using artificial intelligence techniques: Current state and open problems (2012)
3. Bahukhandi, A.: Metastability (2003)
4. Bjorck, N., Gomes, C.P., Selman, B., Weinberger, K.Q.: Understanding batch normalization. In: S. Bengio, H. Wallach, H. Larochelle, K. Grauman, N. Cesa-Bianchi, R. Garnett (eds.) Advances in Neural Information Processing Systems 31, pp. 7694 – 7705. Curran Associates, Inc. (2018). URL http://papers.nips.cc/paper/7996-understanding-batch-normalization.pdf

5. Farid, A.: An overview of game testing techniques (2011)
6. Gudmundsson, S., Eisen, P., Poromaa, E., Nodet, A., Purmonen, S., Kozakowski, B., Meurling, R., Cao, L.: Human-like playtesting with deep learning. pp. 1–8 (2018). DOI 10.1109/CIG.2018.8490442
7. Guzdial, M., Li, B., Riedl, M.O.: Game engine learning from video. In: Proceedings of the Twenty-Sixth International Joint Conference on Artificial Intelligence, IJCAI-17, pp. 3707–3713 (2017). DOI 10.24963/ijcai.2017/518. URL https://doi.org/10.24963/ijcai.2017/518
8. Howard, A., Sandler, M., Chu, G., Chen, L.C., Chen, B., Tan, M., Wang, W., Zhu, Y., Pang, R., Vasudevan, V., Le, Q.V., Adam, H.: Searching for mobilenetv3. ArXiv **abs/1905.02244** (2019)
9. Kamaldinov, I., Makarov, I.: Deep reinforcement learning in match-3 game. In: 2019 IEEE Conference on Games (CoG), pp. 1–4 (2019). DOI 10.1109/CIG.2019.8848003
10. Mnih, V., Kavukcuoglu, K., Silver, D., Graves, A., Antonoglou, I., Wierstra, D., Riedmiller, M.A.: Playing atari with deep reinforcement learning. CoRR **abs/1312.5602** (2013). URL http://arxiv.org/abs/1312.5602
11. Noorian, M., Bagheri, E., Du, W.: Machine learning-based software testing: Towards a classification framework. In: Proceedings of the 23rd International Conference on Software Engineering & Knowledge Engineering (SEKE'2011), Eden Roc Renaissance, Miami Beach, USA, July 7-9, 2011, pp. 225–229. Knowledge Systems Institute Graduate School (2011)
12. Poromaa, E.R.: Crushing candy crush : Predicting human success rate in a mobile game using monte-carlo tree search. Master's thesis, KTH, School of Computer Science and Communication (CSC) (2017)
13. Purmonen, S.: Predicting game level difficulty using deep neural networks. Master's thesis, KTH, School of Computer Science and Communication (CSC) (2017)
14. Sutton, R.S., Barto, A.G.: Reinforcement Learning: An Introduction, second edn. The MIT Press (2018). URL http://incompleteideas.net/book/the-book-2nd.html
15. Walsh, T.: Candy crush is np-hard. ArXiv **abs/1403.1911** (2014)
16. Zhao, Y., Borovikov, I., Beirami, A., Rupert, J., Somers, C., Harder, J., de Mesentier Silva, F., Kolen, J., Pinto, J., Pourabolghasem, R., Chaput, H., Pestrak, J., Sardari, M., Lin, L., Aghdaie, N., Zaman, K.A.: Winning isn't everything: Training human-like agents for playtesting and game AI. CoRR **abs/1903.10545** (2019). URL http://arxiv.org/abs/1903.10545